\documentclass[10pt,prl,aps,showpacs,twocolumn,unsortedaddress]{revtex4-1}
\usepackage{graphicx}
\usepackage{amsmath,amssymb}
\usepackage{bm}
\usepackage{subfigure}
\usepackage{color}
\usepackage{bbold}
\begin{document}
\newcommand{\bra}[1]{\langle #1|}
\newcommand{\ket}[1]{|#1\rangle}
\newcommand{\braket}[2]{\langle #1|#2\rangle}
\newcommand{\HRule}[1]{\rule{\linewidth}{#1}}

\title{
Phase diffusion in a Bose-Einstein condensate of light
}

\author{A.-W. de Leeuw}
\email{A.deLeeuw1@uu.nl}
\author{E.C.I. van der Wurff}
\author{R.A. Duine}
\author{H.T.C. Stoof}
\affiliation{Institute for Theoretical Physics and Center for Extreme Matter and Emergent Phenomena, Utrecht
University, Leuvenlaan 4, 3584 CE Utrecht, The Netherlands}

\date{\today}

\begin{abstract}
We study phase diffusion in a Bose-Einstein condensate of light in a dye-filled optical microcavity, i.e., the spreading of the probability distribution for the condensate phase. To observe this phenomenon, we propose an interference experiment between the condensed photons and an external laser. We determine the average interference patterns, considering quantum and thermal fluctuations as well as dissipative effects due to the dye. Moreover, we show that a representative outcome of individual measurements can be obtained from a stochastic equation for the global phase of the condensate.
\end{abstract}

\maketitle

\begin{figure}[t]
 \centerline{\includegraphics[scale=1.14]{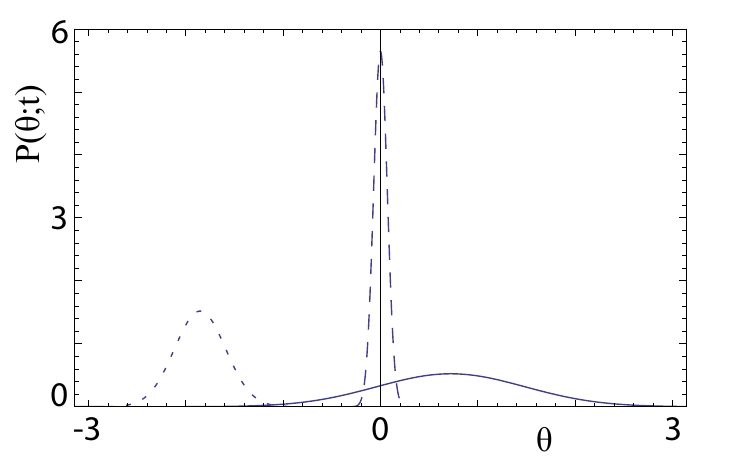}}
 \caption{The probability $P(\theta,t)$ for the Bose-Einstein condensate of photons at different times for $N_{0} = 5 \cdot 10^{4}$ and $\sigma = 10^{-1}$. The dashed, dotted and solid curve are the probability at $t = 0$, $t = t_{\mathrm{col}}$ and $t = 3 t_{\mathrm{col}}$. We clearly see the diffusion of the phase of the condensate if time evolves.} 
 \label{fig:ProbPD2}
\end{figure}

\begin{figure}[t]
 \centerline{\includegraphics[scale=1.14]{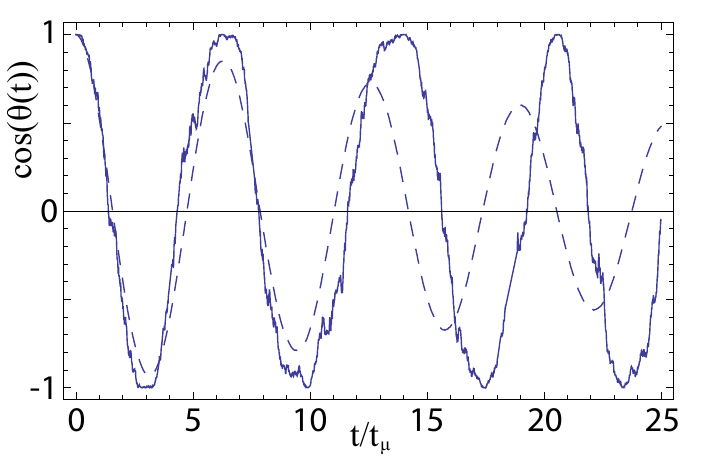}}
 \caption{The result of $\cos(\theta(t))$ as a function of  $t / t_{\mathrm{osc}}$ with $t_{\mathrm{osc}} = \hbar / \mu \simeq 5 \cdot 10^{-10}\,\mathrm{s}$. Here $\theta(t)$ is a solution to the Langevin equation describing the dynamics of the phase of the condensate for $N(t) = \langle N \rangle = 4 \cdot 10^{4}$, $\alpha = 10^{-1}$ and $\hbar \beta \simeq 2.5 \cdot 10^{-14}\,\mathrm{s}$. The solid curve is the result for an arbitrary noise configuration and the dashed curve represents the average over 500 different configurations of the noise.} 
 \label{fig:Simu}
\end{figure}

\textit{Introduction.---} Phase transitions are every-day phenomena that have many high-tech applications in daily life, such as for example the isotropic-nematic phase transition in LCD screens. Additionally, phase transitions are often encountered in fundamental research, such as in the description of superconductivity \cite{SC1} and the electroweak and QCD phase transition in cosmology \cite{Cos1, Cos2, Cos3}. As a result, throughout history much effort has been put in understanding phase transitions. A crucial step was the development of Landau theory in 1937 \cite{Landau}, which provided a general framework to describe symmetry-breaking phase transitions. 
\newline
\indent Many phase transitions are associated with spontaneous symmetry breaking \cite{SSB1, SSB2}. In these transitions the state of the system after the phase transition does not show the same symmetry as the Hamiltonian. As an illustration of spontaneous symmetry breaking, we consider the Heisenberg model for ferromagnetism \cite{Heisenberg1}. In this system the Hamiltonian is invariant under rotations of the spins. However, after undergoing the transition the spins align in a particular direction, and the state of the system breaks spin rotation invariance. However, the original symmetry still has consequences as a global rotation of all spins leaves the energy invariant. Therefore, the ordered phase is infinitely degenerate and spontaneous symmetry breaking by itself does not provide an explanation which particular ground state the system chooses. 
\newline
\indent We can investigate this problem by looking at the probability distribution of the quantum-mechanical observable that acquires a non-zero expectation value upon undergoing the transition. In the context of atomic gases and Bose-Einstein condensation \cite{Bose, Einstein}, the Hamiltonian is invariant under global $U(1)$ transformations associated with the conservation of the number of atoms. Therefore, the number of condensed particles and the phase of the condensate are conjugate variables. Heisenbergs uncertainty principle implies that for a fixed number of condensed particles the phase of the condensate fluctuates. Thus, in finite-sized condensates the phase is not fixed and the system is not in a state with a definite phase. Rather, the phase of the condensate is characterized by a probability distribution, which can have non-trivial dynamics of its own. In Bose-Einstein condensates this phenomenon is known as phase diffusion \cite{PDBEC, PDBEC2}. 
\newline
\indent Considerable theoretical work has been done on phase diffusion in atomic condensates \cite{PDAtom1, PDAtom2, PDAtom3, PDAtom4, PDBEC4}. Experimentally, there also have been some attempts to measure this phenomenon \cite{PDBEC5,PDBEC3}, but up to now there is no experimental evidence of phase diffusion. 
\begin{figure}[t]
 \centerline{\includegraphics[scale=2.5]{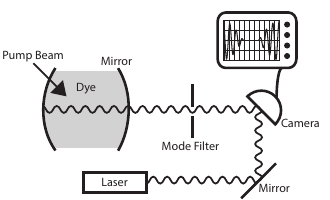}}
 \caption{Proposal for an experimental setup to measure the phase diffusion of the Bose-Einstein condensate of photons in a dye-filled microcavity. The mode filter selects the condensate mode of the light that leaks through the mirror. These condensed photons interfer with an external laser, and by measuring the intensity of the combined signal we obtain information about the phase diffusion of the Bose-Einstein condensate of photons.} 
 \label{fig:ExpSetup}
\end{figure}
More recently, quasiparticle Bose-Einstein condensates, such as condensates of magnons, exciton-polaritons and photons have been observed \cite{BECmagnon, BECpolariton, BECpolariton2, BECphoton}. Since the phase of light can be obtained from a relatively simple interference experiment, the discovery of Bose-Einstein condensation of photons in particular, opens up a new avenue to investigate phase diffusion in Bose-Einstein condensates. 
\newline
\indent In this Letter, we therefore study phase diffusion in a Bose-Einstein condensate of photons. We propose an interference experiment between the condensed photons and an external laser to measure phase diffusion in the photonic condensate. Since phase diffusion is governed by both quantum and thermal fluctuations of the number of condensed particles, we calculate average interference patterns for both cases seperately. Moreover, for the at present experimentally most relevant situation where thermal fluctuations dominate, we show that representative results of individual measurements can be obtained from a stochastic equation for the phase of the condensate.
\newline
\indent \textit{Quantum fluctuations.---} Experimentally, information on the phase of the condensate can be inferred from interfering the electric field of the photon condensate with an external laser and measuring the intensity of the combined signal. We assume that the laser is frequency-locked to the homogeneous non-interacting energy of a condensed photon, and without loss of generality we assume that the distance from the laser and the condensed photons to the detector is the same. A schematic picture of the experimental setup is shown in Fig.\,\ref{fig:ExpSetup}. 
\newline
\indent Since for a finite-size condensate of photons the phase is not well-defined, we introduce a density operator $\hat{\rho}$ that takes into account that the photons can be in a superposition of different coherent states with different phases. Following Ref.\,\cite{QuanOpt}, we write for the intensity of the combined signal of the laser and the condensate at the detector as
\begin{align}
\bar{I}({\bf r},t) = \mathrm{Tr}\left[\hat{\rho} \hat{E}^{-}({\bf r},t)  \hat{E}^{+}({\bf r},t) \right],
\end{align}
where the bar denotes the average and with $\hat{E}^{-}({\bf r},t)$ and $\hat{E}^{+}({\bf r},t)$ respectively the negative and positive frequency part of the sum of the electric field of the laser and the Bose-Einstein condensate.
\newline
\indent For our system the relevant basis states are the coherent states $\ket{\theta_{\mathrm{C}}}\ket{\theta_{\mathrm{L}}}$, where $\ket{\theta_{\mathrm{L}}}$ is a coherent state of the laser with phase $\theta_{\mathrm{L}}$ and $\ket{\theta_{\mathrm{C}}}$ a coherent state of the Bose-Einstein condensate with a certain phase $\theta_{\mathrm{C}}$. In the following, we assume without loss of generality that $\theta_{\mathrm{L}} = 0$. By using properties of these coherent states (see e.g. Ref.\,\cite{Atland}), we obtain for the interference contribution of the intensity
\begin{align}\label{eq:ifp2}
\bar{I}_{\mathrm{I}}({\bf r},t) &:= \bar{I}({\bf r},t) -  I_{\mathrm{L}}({\bf r},t)  - I_{\mathrm{C}}({\bf r},t) \\ \nonumber
&=  2 A_{\mathrm{I}}({\bf r},t) \int_{0}^{2\pi} d\theta\, P({\theta},t) \cos(\theta),
\end{align}
where $I_{\mathrm{L}}({\bf r},t)$ and $I_{\mathrm{C}}({\bf r},t)$ are the intensity of respectively the laser and the condensed photons. Furthermore, $A_{\mathrm{I}}({\bf r},t)$ is a prefactor that is the product of the amplitude of the electric field of the condensed photons and of the external laser. Moreover, $P(\theta,t)$ is the probability for the Bose-Einstein condensate to have a phase $\theta$. Since the intensity of the photons coming from the condensate is independent of the phase, this interference part of the intensity is the only relevant contribution for observing phase diffusion. 
\newline
\indent For an explicit expression of the intensity as a function of time, we need to determine the probability $P(\theta,t)$. In analogy with Ref.\,\cite{UQF}, we obtain this probability by quantizing a field theory that describes the dynamics of the phase of a Bose-Einstein condensate of photons. These photons are equivalent to a two-dimensional harmonically trapped gas of bosons with effective mass $m$ \cite{BECphoton}. Furthermore, they have an effective contact interaction with strength $T^{2B}$ and a constant zero-momentum energy $m c^{2}$, with $c$ the speed of light in the medium. Note that we have assumed the laser to be frequency locked to $m c^{2}$. Therefore, in imaginary time the relevant action is given by
\begin{align}
S[\psi^{*},\psi] &= \int_{0}^{\hbar \beta} d\tau \int d{\bf x} \, \psi^{*}({\bf x}, \tau) \Bigg{(} \hbar \frac{\partial}{\partial \tau} - \frac{\hbar^{2} \nabla^{2}}{2 m} \\ \nonumber
&-\mu + \frac{1}{2} m \Omega^{2} |{\bf x}|^{2} + \frac{T^{2B}}{2} |\psi({\bf x}, \tau)|^{2} \Bigg{)} \psi({\bf x}, \tau),
\end{align}
where $\beta = 1 / k_{\mathrm{B}} T$ with $T$ the temperature, $\mu$ is the chemical potential of the photons with respect to the energy $m c^{2}$ and $\Omega$ is the harmonic trapping frequency. In the following we use numerical values for $\Omega$, $T^{2B}$ and $m$ as given in Ref.\,\cite{BECphoton}.
\newline
\indent To extract the dynamics of the global phase, we substitute $\psi({\bf x}, \tau) = \sqrt{\rho({\bf x},\tau)}e^{i \theta (\tau)}$. Moreover, we consider the Thomas-Fermi limit relevant for experiments and therefore can neglect the gradient of the density profile $\rho({\bf x},\tau)$. By integrating out the density field $\rho({\bf x},\tau)$ and performing a Wick rotation $\tau \rightarrow it$, we find an effective action for the global phase. Quantizing this theory, we find that the wavefunction $\Psi(\theta,t)$ obeys
\begin{align}\label{eq:schrod}
i \hbar \frac{\partial \Psi(\theta,t)}{\partial t} = - D \left(\frac{\partial}{\partial \theta} + i N_{0} \right)^{2} \Psi(\theta,t),
\end{align}
where $N_{0} = \int d{\bf x} \,  \bar{\rho}({\bf x}) $ is the average number of condensed photons, and the diffusion constant is defined as $D = T^{2B} / 2 \pi R_{\mathrm{TF}}^{2}$ with $R_{\mathrm{TF}}$ the Thomas-Fermi radius of the photon condensate. The general solution to this equation reads
\begin{align}\label{eq:wf}
\Psi(\theta,t) = \sum_{n \in \mathbb{Z}} c_{n} \exp\left\{- \frac{i D (n+N_{0})^{2} t}{\hbar} + i n \theta \right\},
\end{align}
where the coefficients $c_{n}$ are determined by the initial condition of the wavefunction.
\newline
\indent In order to demonstrate the phase diffusion and to calculate a typical interference pattern, we consider the example that the initial wavefunction is a superposition of Gaussians centered around $\theta = 0 \, \mathrm{mod} \, 2 \pi$,
\begin{align}
\Psi(\theta,0) = \frac{1}{(\pi \sigma^{2})^{1/4}} \sum_{n \in \mathbb{Z}} \exp\left\{- \frac{(\theta + 2 \pi n)^{2}}{2 \sigma^{2}} \right\}.
\end{align}
Taking this superposition ensures that the wavefunction is periodic, i.e., $\Psi(\theta,0) = \Psi(\theta + 2 \pi,0)$. In principle we have a slightly different normalization factor, but for the small values of $\sigma < 1$ considered here, this is a very good approximation. In experiments one would measure the phase of the condensate and then look at its dynamics. Hence, we start from a wavefunction that is strongly peaked and therefore we can use in good approximation that $\sigma < 1$.
\newline
\indent For this initial wavefunction, we can determine $c_{n}$ exactly and obtain an analytic expression for the probability $P(\theta,t) = |\Psi(\theta,t)|^{2}$. Typical plots of this probability are shown in Fig.\,\ref{fig:ProbPD2}. 
At $t = 0$ we have a sharp peak and therefore the phase of the condensate is well-defined. However, if time evolves the peak smears out and moves its position linearly with time. As time evolves even further, the probability again regains its original shape. This phenomenon is known as collapse and revival of the wavefunction, and is a consequence of the invariance of the wavefunction for $t \rightarrow t + 2 \pi k \hbar / D$ for every integer $k$, as can be deduced from Eq.\,\eqref{eq:wf}. Hereafter, cycles of collapse and revival of the wavefunction occur.
\newline
\indent Moreover, we use our expressions for the probability to obtain the average interference pattern as defined in Eq.\,\eqref{eq:ifp2}. Again for small $\sigma < 1$, we find
\begin{align}\label{eq:intsigma}
\bar{I}&_{\mathrm{I}}({\bf r},t) = \frac{2 \sigma A_{\mathrm{I}}({\bf r},t)}{\sqrt{\pi}}  \cos\left(\frac{5 (1 + 2 N_{0}) \sigma t}{2 t_{\mathrm{col}}}\right) \\ \nonumber
&\times \sum_{n \in \mathbb{Z}} \exp\left\{-n (n + 1) \sigma^{2} \right\} \cos\left(\frac{5 n \sigma t}{t_{\mathrm{col}}} \right),
\end{align}
with $t_{\mathrm{col}} = 5 \hbar \sigma / 2 D$. This time gives a measure of the time needed for this pattern to vanish for the first time. Furthermore, this expression contains two other important time scales. The first scale is the oscillation time of the intereference pattern, which for a relatively large number of condensed photons $N_{0} \gg 1$ is given by  $t_{\mathrm{osc}} = \hbar / 2 D N_{0}$. Physically this corresponds to $\hbar / \mu$, with $\mu$ the chemical potential of the condensate. Note that this calculation gives a factor of two difference because of the quadratic expansion of the grand canonical energy. The second time scale, given by $t_{\mathrm{rev}} = 2 \pi \hbar / D$, is the revival time for which the interference pattern returns to its original shape. Note that this time scale is larger than $t_{\mathrm{osc}}$ by a factor $4 \pi N_{0}$. Furthermore, in the thermodynamic limit $N_{0} \rightarrow \infty$, we find that $D \propto 1 / \sqrt{N_{0}} \rightarrow 0$ and both $t_{\mathrm{col}} \rightarrow \infty$ and $t_{\mathrm{rev}} \rightarrow \infty$. Hence, in the thermodynamic limit the condensate can be described as a symmetry-broken phase. 
\newline
\indent In the previous calculations we ignored that the photons are in a dye-filled optical microcavity, and that there is dissipation through the interaction with these dye molecules. As is shown in Ref.\,\cite{AW}, for low energies these interaction effects can in very good approximation be represented by one single dimensionless damping parameter $\alpha$. To incorporate this damping into our calculation, we note that damping results into finite lifetimes for states with a non-zero energy. Therefore, as a first attempt to include dissipation, we change Eq.\,\eqref{eq:wf} into
\begin{align}
\Psi(\theta,t) = \sum_{n \in \mathbb{Z}} c_{n} \int dE\, \rho(E,n) \exp\left\{- \frac{i E t}{\hbar} + i n \theta \right\},
\end{align}
where the spectral function $\rho(E,n)$ is given by
\begin{align}
\rho(E,n) = \frac{1}{\pi} \frac{\alpha E}{(E - D (n + N_{0})^{2})^{2} + \alpha^{2} E^{2}}.
\end{align}
A consequence of approximating the dissipation effects with its low-energy limit is a violation of the sum rule, since the integral of the spectral function over all energies gives $1/(1+ \alpha^{2})$. However, the experimental value of $\alpha$ is rather small and therefore this approximation only leads to a small deviation. 
\newline
\indent For a relatively small number of condensed photons, the interference pattern with dissipation reads
\begin{align}\label{eq:Ialpha}
\bar{I}({\bf r},t; \alpha) \simeq e^{-t/t_{\mathrm{dis}}} \bar{I}_{\mathrm{I}}({\bf r},t),
\end{align}
where $t_{\mathrm{dis}} = \hbar/ 4 \alpha D N_{0}^{2}$ and $\bar{I}_{\mathrm{I}}({\bf r},t)$ is given by Eq.\,\eqref{eq:intsigma}. Thus with dissipation there is another time scale $t_{\mathrm{dis}}$, which indicates the decay time of the interference pattern. For very large condensates $N_{0} \gg 1$, the low-energy approximation of the dissipation is no longer valid and we have to incorporate the complete energy dependence of the photon decay rate $\Gamma(E)$ as calculated in Ref.\,\cite{AW}. In good approximation the dissipation time scale is then found by replacing $\alpha D N_{0}^{2}$ by $\hbar \Gamma(D N_{0}^{2})/2$.
\newline
\indent \textit{Thermal fluctuations.---} Analogously with Ref.\,\cite{Keldysh}, we describe the thermal fluctuations with a Langevin field equation. As mentioned before, we incorporate the interaction with the molecules by one dimensionless parameter $\alpha$. Furthermore, we neglect the effects of the non-condensed photons. By following the lines of Ref.\,\cite{Stoch}, we separate the dynamics of the number of photons $N(t)$ and their global phase $\theta(t)$, and find
\begin{align}\label{eq:StochPhase}
(1 + \alpha^{2}) \hbar \dot{\theta}(t) &= -\mu + \sqrt{\frac{1 + \alpha^{2}}{N(t)}} \nu(t), \\ \nonumber
(1 + \alpha^{2}) \dot{N}(t) &= -\frac{2 \alpha \mu}{\hbar} N(t) + 2 \sqrt{N(t) (1 + \alpha^{2})} \eta(t),
\end{align}
where the stochastic generalized forces $\eta(t)$ and $\nu(t)$ are Gaussian and obey
\begin{align}
\langle \nu(t) \rangle &= \langle \eta(t) \rangle = \langle \eta(t) \nu(t^{\prime}) \rangle = 0, \\ \nonumber
\langle \nu(t) \nu(t^{\prime}) \rangle &= \hbar^{2} \langle \eta(t) \eta(t^{\prime}) \rangle \simeq \frac{\alpha \hbar}{\beta} \delta(t - t^{\prime}).
\end{align}
Since we are dealing with Bose-Einstein condensation, we used the fluctuation-dissipation theorem for large occupation numbers. Because the photons are at room temperature, we expect this to be a very good approximation. Furthermore, we note that the strength of the noise for the number $N(t)$ and phase $\theta(t)$ of the condensed photons scales differently with the number of condensed photons. For larger number of photons the fluctuations in the particle number increase, but the fluctuations in the global phase decrease. Moreover, in the thermodynamic limit the noise for the global phase vanishes, and we obtain again a condensate with a well-defined phase.
\newline
\indent As the description of the thermal fluctuations is different from the quantum fluctuations, we need to modify our expression for the interference pattern. In the previous section we found an expression by taking the average over an ensemble consisting of various quantum states, each with a certain probability. A single experimental measurement, however, typically yields
\begin{align}\label{eq:ifp}
I_{\mathrm{I}}({\bf r},t) =  2 A_{\mathrm{I}}({\bf r},t) \cos(\theta(t)),
\end{align}
where $\theta(t)$ is the solution of Eqs.\,\eqref{eq:StochPhase} for one realization of the noise. As mentioned before, the fluctuations of the phase of the condensate are only present in this interference part of the intensity, and therefore we are primarly interested in this part of the intensity. Moreover, to highlight the fluctuations of the phase we would like to minimize the fluctuations in the intensity of the external laser and the light coming from the condensate. Since the intensity of the condensate is proportional to the number of condensed photons, we are interested in the regime with small number fluctuations. As can be deduced from the experimental results in Ref.\,\cite{NumFluc} and Eqs.\,\eqref{eq:StochPhase}, the number fluctuations decrease for increasing condensate fractions. Therefore, we consider large condensate fractions such that the fluctuations in the interference pattern are dominated by phase fluctuations, and we take $N(t) = \langle N(t) \rangle : = N_{0}$.
\newline
\indent
In Fig.\,\ref{fig:Simu} we show the result for $\cos(\theta(t))$, where $\theta(t)$ is a solution to the stochastic Eqs.\,\eqref{eq:StochPhase} for a condensate fraction of roughly $35\,\%$. The solid curve gives the interference pattern for a certain realization of the stochastic forces. Every realization of the noise results into a different interference pattern, and therefore every individual measurement will give a different result. 
However, once we average over more and more noise realizations $\langle \cos(\theta(t)) \rangle$ converges, and we do observe the decay associated with the dissipation.
\newline
\indent In order to get more information about this decay of the intensity $I_{\mathrm{I}}({\bf r},t)$, we have to take the average of Eq.\,\eqref{eq:ifp} over all noise configurations. By using the Fokker-Planck equation as derived in Ref.\,\cite{Stoch}, we find
\begin{align}
(1+\alpha^{2})\hbar \frac{\partial}{\partial t} \langle \cos(\theta) \rangle &= \mu \langle \sin(\theta) \rangle - \frac{\alpha}{2 \beta N_{0}} \langle \cos(\theta) \rangle, \\ \nonumber
(1+\alpha^{2})\hbar \frac{\partial}{\partial t} \langle \sin(\theta) \rangle &= -\mu \langle \cos(\theta) \rangle - \frac{\alpha}{2 \beta N_{0}} \langle \sin(\theta) \rangle,
\end{align}
These equations admit analytic solutions and, by neglecting contributions of order $\alpha^{2}$, we find for the average of the interference part of the intensity
\begin{align}\label{eq:IT}
\langle I_{\mathrm{I}}({\bf r},t) \rangle = 2 A_{\mathrm{I}}({\bf r},t) \exp\left\{-\frac{\alpha t}{2 \hbar \beta N_{0}}\right\} \cos\left(\frac{\mu t}{\hbar}\right),
\end{align}
which coincides with the result in Fig.\,\ref{fig:Simu} where we averaged over 500 noise realizations.
\newline
\indent \textit{Discussion and conclusion.---} In the previous sections, we gave a discussion on phase diffusion governed by quantum and thermal fluctuations. Since thermal fluctuations are dominant for the current experiment, there are two important time scales $t_{\mathrm{osc}} = \hbar/\mu$ and $t_{\mathrm{dis}} = 2 \hbar\beta N_{0}/\alpha$. For typical values for the trap frequencies $\Omega$, we obtain that $t_{\mathrm{osc}}$ is in the order of picoseconds. Since this is rather small, we expect that it is challenging to measure these oscillations experimentally. However, for large condensate numbers $N_{0} \gg 1$ and $\alpha$ ranging from $10^{-1}$ to $10^{-2}$ the decay time $t_{\mathrm{dis}}$ is in the nanoseconds regime, which is within the precision of current devices.
\newline
\indent In conclusion, we have calculated the phase diffusion of a Bose-Einstein condensate of photons. We propose an interference experiment of the condensed photons with an external laser to observe this phase diffusion experimentally. Furthermore, we have shown that the typical outcome of individual experiments can be obtained from a stochastic equation for the phase of the condensate. Finally, we have demonstrated that thermal fluctuations dominate, and we obtained that the decay time of the average interference pattern is in the nanosecond regime, which is a accessible time scale in experiments.
\newline
\indent Although the calculations in this work are specific for a Bose-Einstein condensate of photons, the concepts and ideas presented in this paper are also applicable to Bose-Einstein condensation of exciton-polaritons. Namely, also in these Bose-Einstein condensates we can get experimental information about the global phase of the condensate. For example, in Refs.\,\cite{ExPol1, ExPol2} the relative global phase of two coupled exciton-polariton condensates is measured in order to investigate Josephson oscillations. Therefore, it is worthwile to apply the presented theory to Bose-Einstein condensation of exciton-polaritons.
\newline
\indent It is a pleasure to thank Dries van Oosten for useful discussions. This work was supported by the Stichting voor Fundamenteel Onderzoek der Materie (FOM) and the European Research Council (ERC), and is part of the D-ITP Consortium, a program of the Netherlands Organisation for Scientific Research (NWO) that is funded by the Dutch Ministry of Education, Culture and Science (OCW).

\end{document}